**Changes in seismicity pattern due to the 2016 Kumamoto earthquake sequence and implications for improving the foreshock traffic-light system**

K. Z. Nanjo[1,2,3,*], J. Izutsu[4], Y. Orihara[5], M. Kamogawa[1]

[1]Global Center for Asian and Regional Research, University of Shizuoka, 3-6-1, Takajo, Aoi-ku, Shizuoka 420-0839, Japan.

Tel: +81-54-245-5600 Fax: +81-54-245-5603

[2]Center for Integrated Research and Education of Natural Hazards, Shizuoka University, 836, Oya, Suruga-ku, Shizuoka 422-8529, Japan.

[3]Institute of Statistical Mathematics, 10-3, Midori-cho, Tachikawa, Tokyo 190-8562, Japan.

[4]International Digital Earth Applied Science Research Center, Chubu University, 1200, Matsumoto-cho, Kasugai, Aichi 487-8501, Japan.

[5]Tokyo Gakugei University, 4-1-1, Nukuikita-machi, Koganei, Tokyo 184-8501, Japan.

*Correspondence to: nanjo@u-shizuoka-ken.ac.jp

**Highlights**

- Changes in seismicity and stress distribution due to the 2016 Kumamoto earthquakes

- Stress field revealed aseismic pre- and after-slow slips along the causative faults

- System to discriminate foreshocks and aftershocks was tested for the Kumamoto case

- Test showed that seismicity before the largest event was regarded as aftershocks

- Aseismic slips and changes in stress are keys to modify the discrimination system

**Abstract**

Crustal deformation due to the 2016 earthquake sequence in Kumamoto, Japan, that culminated in a preceding earthquake of magnitude $M$6.5 and a subsequent $M$7.3 earthquake 28 hours later, caused stress perturbation on and around the causative Futagawa-Hinagu fault zone. Monitoring changes in seismicity pattern along this zone plays a role in understanding the process before and after major earthquakes. For this purpose, stress-dependent laws in statistical seismology can be used: the Gutenberg-Richter frequency-size law and the Omori-Utsu aftershock-decay law. We review the results obtained by using these laws in previous studies to show a zone of high stress near the eventual epicenters of the $M$6.5 and $M$7.3 earthquakes before the start of the Kumamoto sequence, and after it, showing a decreasing trend in stress along the Futagawa-Hinagu fault zone. Detailed analysis suggests aseismic slips along the causative fault zone. The aseismic preslip locally reduced stress just prior to the $M$7.3 earthquake near its epicenter. Recently, a system was proposed by Gulia and Wiemer (2019) that utilizes the Gutenberg-Richter frequency-size law to judge, immediately after a large earthquake, whether it was the mainshock or a foreshock to a future event. Based on the reviewed results and our new results, further research that takes into account the spatial variation of frequency-size distribution, allowing the exploration of the possibility of a local preslip of a future nearby earthquake, is needed to improve this system.

1. **Introduction**

The 2016 Kumamoto earthquake sequence (Fig. 1) that caused serious damage to Kumamoto, in central Kyushu, Japan (Hashimoto et al., 2017), began with an earthquake of magnitude ($M$) 6.5 on April 14, 2016, at 21:26, on the northern section of the Hinagu fault zone (FZ). It was followed by numerous events, including a $M$6.4 earthquake on April 15, at 00:03. These $M$6.5-class earthquakes, which were considered as foreshocks, involved a predominantly right-lateral strike slip. Eventually, on April 16, at 01:25, a $M$7.3 mainshock occurred on the Futagawa FZ that conjugated with the northern section of the Hinagu FZ, revealing a similar mechanism of fault motion as the $M$6.5-class foreshocks. The $M$7.3 mainshock started at the westernmost part of the Futagawa FZ, and its rupture propagated toward the east and terminated around Mt. Aso (Yagi et al., 2016). The Earthquake Research Committee (ERC) concluded that the Futagawa FZ and the northern section of the Hinagu FZ had been activated (ERC, 2016a). Over the first three years since the $M$7.3 mainshock (Fig. 1b), more than 27,000 aftershocks with $M \geq 1$ occurred, including more than 100 events with $M \geq 4$. The broader context of the Kumamoto earthquakes is that the Futagawa-Hinagu FZ is encompassed by the Beppu–Shimabara graben, a geological formation that runs across the middle of Kyushu, from Beppu Bay in the east to the Shimabara Peninsula in the west (Inset of Fig. 1). This is understood as being the result of crustal deformation caused by the rifting and spreading of the Okinawa Trough, which is viewed as a continuation of the Beppu–Shimabara graben (Tada 1985).

Monitoring crustal deformation and seismicity along the causative faults plays a role in understanding the process before and after major earthquakes. Numerical simulation and rock mechanics experiments suggested the occurrence of pre- and post-seismic slip transients. Some studies implied long-term precursory slip acceleration while other studies indicated a nucleation process that initially involves stable and slow rupture growth (Hori and Miyazaki, 2010; Yoshida and Kato, 2003; Dieterich, 1979; Lapusta et al., 2000). However, it is generally accepted that a precursory slip before a large earthquake is not always observed. This may be because this phenomenon does not always emerge in nature and/or due to difficulty in observing a small slip. On the other hand, a post-seismic slip was observed for many earthquakes. In the aftermath of the Kumamoto earthquakes, slow deformation caused by a post-seismic slip of the causative Futagawa-Hinagu FZ, together with flow in the Earth's interior, has been detected by data from the GEONET: GNSS Earth Observation Network System (Moore et al., 2017; Pollitz et al., 2017).

Recognizing earthquakes as foreshocks in real time would provide a valuable forecasting capability. However, decades of work have failed to establish a robust feature of individual foreshocks that distinguishes them from mainshocks or their aftershocks (Dascher-Cousineau et al., 2020). Foreshocks may be distinctive due to some precursory loading process or influence from the locked zone of the subsequent mainshock, neither of which will exist in the aftershock sequence (Brodsky and Lay, 2014). A fundamental question regarding the Kumamoto sequence is whether or not any indication of an impending large earthquake had appeared before the $M$7.3 mainshock. The Japan Meteorological Agency (JMA) held a press conference after the $M$6.5 earthquake to warn of the possibility of large aftershocks with further damage. However, the JMA did not considered the occurrence of large earthquakes, according to the ERC (1998) protocol (see also ERC, 2016b), so no information was made public on the increased probability of $M$7 or larger earthquakes.

In a recent study, Gulia and Wiemer (2019) proposed a Foreshock Traffic-Light System (FTLS) that relies on abrupt changes in $b$-values of the Gutenberg-Richter (GR) law (Gutenberg and Richter, 1944), relative to background values. The GR law is given as $\log_{10}N = a-bM$, where $a$ and $b$ are constants, and $N$ is the cumulative number of earthquakes with a magnitude larger than or equal to $M$. The approach utilizes high-resolution earthquake catalogs to monitor localized regions around the largest earthquakes and to distinguish foreshock sequences (reduced $b$-values) from aftershock sequences (increased $b$-values) (Gulia et al., 2020; Dascher-Cousineau et al., 2020). Evidence supporting the FTLS hypothesis emerged from investigating a timeseries of two sequences: the Norcia sequence in Italy and the Kumamoto sequence (Gulia and Wiemer, 2019). Their results of the Kumamoto case showed that $b$-values in the time interval between the $M$6.5 and $M$7.3 events first dropped from the background level and then increased over time, and that the $b$-values at the intermediate and ending times were below the background level, triggering a red traffic light (Table 1 of Gulia and Wiemer, 2019) and suggesting that seismicity in between was not considered as an aftershock sequence, rather as a foreshock one. Once the $M$7.3 mainshock occurred, $b$-values of the subsequent events increased strongly by 20-40%, triggering a green traffic light, suggesting that these events were aftershocks.

We first review our previous studies (Nanjo et al., 2016, 2019; Nanjo and Yoshida, 2017) that used the $b$-value of the GR law and the $p$-value of the Omori-Utsu (OU) law (Omori, 1894; Utsu, 1961) to explore the changes in seismicity before and after the Kumamoto mainshock. The OU law is given as $\lambda = k(c + t)^{-p}$, where $t$ is the time since the occurrence of a mainshock, $\lambda$ is the number of aftershocks per unit time at $t$ with $M$ greater than or equal to a cutoff magnitude, and $c$, $k$, and $p$ are constants. The $b$-value is inversely proportional to differential stress (Scholz, 1968; 2015), and the $p$-value can be used to infer stressing history (Dieterich, 1994). The main conclusion is that local- and time-dependent variation in $b$- and $p$-values was linked with changes in stress state associated with aseismic slips and a coseismic one along the causative Futagawa-Hinagu FZ. Based on this review, we revisited the applicability of FTLS to the Kumamoto case and discuss the possibility of improving it.

2. **A review of studies on $b$- and $p$-values associated with the Kumamoto earthquakes**

Understanding the stress state is of key importance for earthquake forecasting and hazard assessment. However, it is rather difficult to measure stress directly underground. Statistical parameters of seismicity might provide a useful way to monitor stress evolution on faults. Therefore, monitoring changes in seismicity pattern along a FZ plays an important role in understanding the process before and after major earthquakes. A set of three papers (Nanjo et a., 2016, 2019; Nanjo and Yoshida, 201) conducted a systematic study on the seismicity associated with the entire process of the Kumamoto sequence.

Using earthquake data since 2000 from the JMA catalog, Nanjo et al. (2016) showed that seismicity before the start of the Kumamoto sequence revealed a zone of low $b$-values near the eventual epicenters of the $M$7.3 mainshock and two $M$6.5-class foreshocks (Fig. 2a). The precursory duration to the $M$7.3 mainshock was found to be unknown based on observation of an approximately constant $b$-value over the past 16 years (Nanjo et al., 2016). A decrease in $b$ tracking stress buildup, as expected from a laboratory experiment (Scholz, 1968; Lei, 2003), might not be observable for the decade-scale observation of active faults, because it was too short to observe such a decrease in $b$ for the Kumamoto case. The mechanism of stress buildup within a FZ is uncertain, but one hypothesis is that a steady slip of the deeper continuation of faults that does not produce earthquakes, but still involves motions across the fault, forces the upper crust around the faults to deform, and thus concentrate stress. The stress field remains difficult to measure directly (Narteau et al., 2009), but it would be helpful if some indirect evidence exists, such as stress inversion by a focal mechanism, or cases of stress buildup. A specific case includes thermally activated creep on major strike-slip faults such as the San Andreas Fault in California (Turcotte et al., 1980; Turcotte and Schubert, 2002). These faults extend deep into the lithosphere, where they are likely to behave plastically, displaying a steady-state creep in deep fault zones.

Assuming that the $M$6.5 foreshock was a mainshock, Nanjo and Yoshida (2017) showed that the OU law with $p > 1$ was correlated with seismic activity for the entire area during the period of 1.16 days between the $M$6.5 and $M$7.3 earthquakes. The same analysis was conducted to obtain $p > 1$ for the northern area (above 32.72°N) and $p < 1$ for the southern area (below 32.72°N), indicating rapid stress relaxation in the northern area, rather than in the southern one. The $b$-values significantly increased over time for the northern area (Fig. 2b), supported by the Utsu test (Utsu, 1999; Schorlemmer et al., 2004). An increase in $b$ over time was observed for the southern area, but this increase was insignificant. The authors suggested that stress around the Futagawa FZ had begun to decrease preceding the $M$7.3 mainshock because they found a large $p$-value ($p > 1$) and a time-dependent increase in $b$ for the northern area.

Nanjo and Yoshida (2017) also pointed out that earthquakes occurred near the hypocenter of the $M$7.3 mainshock in addition to most seismicity around the northern section of the Hinagu FZ. These events occurred later (0.98-0 days preceding the $M$7.3 mainshock). One explanation, based on Onaka (1993), may be that they were produced by a nucleation process of the $M$7.3 mainshock. This process is considered to be promoted by the gradual propagation of an aseismic slip (afterslip) toward the $M$7.3 hypocenter on and around the Hinagu FZ, interpreted from a detailed observation of seismicity migration starting immediately after the occurrence of the $M$6.5 earthquake and terminating at early times after this earthquake (Kato et al., 2016). The finding of the occurrence of earthquakes close to the $M$7.3 hypocenter at later times after the $M$6.5 earthquake implied the growth of a quasi-static preslip before the $M$7.3 mainshock. This preslip might have relaxed stress around the Futagawa FZ that hosted the forthcoming $M$7.3 mainshock. This stress relaxation provided feedback, causing a significantly larger $p$-value than $p = 1$ and a significant increase in the $b$-value (Fig. 2b) in the northern area, including a part of the Futagawa FZ. This relaxation was also inferred from the result of a stress inversion analysis performed by the National Research Institute for Earth Science and Disaster Resilience (2016) that suggested the reduction in normal stress at the Futagawa FZ, i.e., a decrease in compressive stress on the fault. It was considered that afterslip of the $M$6.5 earthquake was not the main cause of the observed change in seismicity, but might set up the conditions that allowed this change in seismicity. This is because a detailed observation (Kato et al., 2016) showed that the afterslip occurred only on and around the Hinagu FZ while the major slip was limited to an early period after the $M$6.5 event.

Nanjo et al. (2019) considered the sequence after the $M$7.3 mainshock until March 2019 to show an area of low $b$-values at the southern end of the large ($M \geq 5$) earthquakes during the period from the start of the Kumamoto sequence (Fig. 2c). To obtain this result of *b*-values, the period from immediately after the *M*7.3 mainshock to the end of 2016 was not considered to remove the effect of the main aftershock sequence that revealed strong temporal variability in *b*. A comparison between the *b* values before the start of the Kumamoto sequence (Fig. 2a) and those on and after 2017 (Fig. 2c) showed a significant increase in *b*-values in most areas along the Futagawa-Hinagu FZ, expected by a release of stress. However, there was an exceptional area showing a decrease in *b*, which coincided with the area of low *b*-values (Fig. 2c). This was interpreted as an indication of overall stress release with a local region of increasing stress. The OU law was found to be applicable to the post-*M*7.3-quake sequence. This law assumed $p > 1$ (fast decay) along the Futagawa FZ and the northern section of the Hinagu FZ and $p < 1$ (slow decay) along the southern section of the Hinagu FZ, with intermediate values ($p \sim 1$) near the boundary between the northern and southern sections of the Hinagu FZ. This boundary corresponded to the area of low and decreasing *b*-values.

      The observation of Nanjo et al. (2019) was explained by a postseismic deformation model with afterslip on the causative faults and viscoelastic relaxation associated with the Beppu-Shimabara graben (Pollitz et al., 2017). Afterslip on the Futagawa FZ and the northern section of the Hinagu FZ, causing a rapid stress decrease, explained large *p*-values ($p > 1$) and an increase in *b*-value. Viscoelastic relaxation of the lower crust and mantle that dominated at a greater distance where the southern section of the Hinagu FZ was involved, caused a gradual decrease in

stress, resulting in small *p*-values ($p < 1$) and again an increase in *b*-values. The afterslip on the northern section of the Hinagu FZ, in contrast to the lack of aftersip on the southern section of the same FZ, might lead to a constant or even an increase in local stress at the boundary between the northern and southern sections. This explained a decrease in *b*-values at this boundary, and also explained an intermediate *p*-value ($p \sim 1$) there, taking into consideration a model for seismicity rate resulting from stressing history (Fig. 8 of Dieterich, 1994).

3.  **Method and Data**

To estimate *b*-values consistently over space and time (Figures 3 and 4), we employed the EMR (entire-magnitude range) technique (Woessner and Wiemer, 2005), which also simultaneously calculates the completeness magnitude $M_c$, above which all events can be detected by a seismic network. EMR applies the maximum-likelihood method to compute the *b*-value to events with a magnitude above $M_c$. Uncertainties in *b* were computed by bootstrapping (Schorlemmer et al., 2003). If log$P_b$, the logarithm of the probability that the *b*-values are not different, is equal to or smaller than -1.3 (log$P_b \leq -1.3$), then the change in *b* is significant (Utsu, 1999; Schorlemmer et al. 2004). Spatiotemporal changes in *b* are known to reflect a state of stress in the Earth's crust (Narteau et al., 2009; Schorlemmer et al., 2005; Scholz, 2015). Patches with small *b*-values on active faults have been observed to coincide with locations of subsequent large earthquakes (Nanjo, 2020;

Schorlemmer and Wiemer, 2005; Tormann et al., 2013; Wang et al., 2021). Investigations using *b*-values have been applied to locate asperities and to estimate frictional properties on the plate interface along subduction zones (e.g. Ghosh et al., 2008; Nanjo et al., 2012; Nanjo and Yoshida, 2018; Sobiesiak et al., 2007; Schurr et al., 2014; Tormann et al., 2015). The results of laboratory experiments indicate a systematic decrease in the *b*-value approaching the time of the entire fracture (Lei, 2003; Scholz, 1968). Foreshocks have been known to show small *b*-values (Gulia and Wiemer, 2019; Gulia et al., 2020; Dascher-Cousineau et al., 2020).

Our dataset was the earthquake catalog maintained by the JMA. We used about $10^6$ earthquakes having $M \geq 0$ during the period from January 1, 2000 to March 11, 2019 with depths shallower than 25 km within the study region (Fig. 1). This period is the same as that considered for a previous paper (Nanjo et al., 2019), and subsets of that period in two other papers (Nanjo et al., 2016; Nanjo and Yoshida, 2017). A *b*-value analysis is critically dependent on a robust estimate of completeness of the processed earthquake data. In particular, underestimates of $M_c$ lead to systematic underestimates of *b*-values. We always paid attention to $M_c$ when assessing $M_c$ at each node and time window. As shown in other studies using the JMA catalog (Nanjo et al., 2010; Schorlemmer et al., 2018), typical values for $M_c$ are presently 0.5~1 in the Kyushu district that includes the Kumamoto region. Small earthquakes in the early stage after a large event are, in many cases, missing from the earthquake catalogs, as they are "masked" by the coda of the large event and overlapped with each other on the seismograms. According to Nanjo and Yoshida (2017) who showed time-dependent decrease of $M_c$ during the period between the $M$6.5 and $M$7.3 quakes, we examined a $M_c$-time pattern (Fig. 4).

## 4. Results and Discussion

A review of previous studies (Nanjo et al., 2016, 2019; Nanjo and Yoshida, 2017) indicates a systematic investigation into the seismicity before, during, and after the Kumamoto earthquakes that started in April of 2016, by analyzing the $b$- and $p$-values (section 2). A brief summary is as follows: (1) the focal areas of the $M \geq 6$ earthquakes were more stressed than others before the start of the Kumamoto sequence; thereafter, stress generally decreased along the causative faults, as inferred from the spatial distribution of $b$- and $p$-values. (2) By using a more detailed analysis, the two statistical parameters suggested aseismic preslip and afterslip along the causative faults. The preslip locally reduced stress just prior to the largest earthquake near its epicenter.

Using the idea of the FTLS, analysis of $b$-values for seismicity during the time interval between the $M$6.5 and $M$7.3 earthquakes (Fig. 2b), in contrast to $b$-values for seismicity before the $M$6.5 earthquake (Fig. 2a), implied a yellow FTLS status for the southern area that included a part of the Hinagu FZ and a green FTLS status for the northern area that included a part of the Futagawa FZ hosting the eventual $M$7.3 earthquake. Here, we examined the inconsistency with the result obtained by Gulia and Wiemer (2019), who showed that a red traffic light was triggered before the *M*7.3 mainshock for the purpose of improving the FTLS.

Fig. 2 shows map views of *b*-values based on seismicity (a) before the *M*6.5 events, (b) during the period between the *M*6.5 and *M*7.3 events (top panel: period 1.16-0.98 days prior to the *M*7.3 events; bottom panel: period 0.98-0 days), and (c) after the *M*7.3 events, reproduced from Nanjo et al. (2016, 2019) and Nanjo and Yoshida (2017). The mapping procedures were the same as those for these studies, but using a unified color code among the maps of *b*-values for ease of visual comparison. The inset of Fig. 2b is a map of log $P_b$, the logarithm of the probability that the *b* values for the two periods, 1.16-0.98 days (top panel) and 0.98-0 days (bottom panel) prior to the *M*7.3 quake are not different, again a reproduction from Nanjo and Yoshida (2017). A general trend of the time-dependent increase in *b* in the interval between the *M*6.5 and *M*7.3 events, which had been shown by Gulia and Wiemer (2019), was also found in Fig. 2b. However, if we considered the southern and northern areas separately, our analysis of *b*-values confirmed a yellow FTLS status for the former area and a green FTLS status for the latter area. For the northern area, which includes a part of the Futagawa FZ on which the impending *M*7.3 mainshock occurred, *b*-values showed an increase over time, to values exceeding 1.0 (bottom panel of Fig. 2b), significantly larger than the background levels (*b* = 0.81~0.83 ± 0.05~0.07) found by Gulia and Wiemer (2019) and those (*b* = 0.6~0.8) in the area of the *M*6.5 and *M*7.3 epicenters (Fig. 2a). On the other hand, for the southern area, which includes a part of the northern section of the Hinagu FZ on which the $M$6.5 foreshock had already occurred, $b$-values were similar to the background level.

The earthquake samples for $b$-value estimation in a grid search (Fig. 2) might contain events from different faults, which might cause possible bias as the $b$-value varies in different fault regimes. The $M$7.3 mainshock and two $M$6.5-class foreshocks belong to two different fault systems. To eliminate doubts about this bias, we conducted a $b$-value timeseries analysis (Fig. 3) based on events in two regions that do not overlap with each other: one region (blue) includes only a part of the Hinagu FZ, and the other (red) includes only a part of the Futagawa FZ. In the latter region, an area close to the $M$7.3 mainshock epicenter was not included, because it was difficult to distinguish between events from the Hinagu FZ and events from the Futagawa FZ. Fig. 4 shows the time dependence of $M_c$, confirming Nanjo and Yoshida (2017), where $M_c$ was simultaneously calculated when applying the EMR method for obtaining the $b$-value. In creating Figs. 3 and 4, we used a moving window approach, whereby the window covered 80 events, and plotted $b$ and $M_c$ at the end of the moving window that they represent. For both regions (Fig. 3), the $b$-values were typically 0.6~0.8 before the $M$6.5 earthquake, where these values have been relatively stable since 2015, as shown in Fig. 3c. The $b$-values showed a rapid decrease just after the $M$6.5 earthquake, to values below 0.5. A transition to increase in $b$ occurred after this decrease. The $b$-values at 0.3 to 0 days relative to the $M$7.3 mainshock were above the background pre-$M$6.5-quake levels for the red region, while those at 0.8 to 0 days were around $b = 0.8$ for the blue region, not yet considered higher than the background values. The difference between the change in $b$-values for the region, including a part of the Futagawa FZ, and for that including a part of the Hinagu FZ is clear. The result supports our implication of a yellow FTLS status for the Hinagu FZ and a green FTLS status for the Futagawa FZ.

Our results for the northern area (indicative of a green FTLS status) suggested a scenario of a reduction in stress induced by growth of aseismic preslip on the Futagawa FZ, as discussed in section 2 (also see Nanjo and Yoshida, 2017), subsequently leading to a much larger chance of a stronger $M$7.3 event. The results by Gulia and Wiemer (2019) (indicative of a red FTLS status) suggested an alternative scenario in which stress on the Futagawa FZ increased, leading to a subsequently much larger chance of a stronger $M$7.3 event. The discrepancy between our study and the Gulia and Wiemer (2019) study is that the former study investigated spatial variation of $b$-values, allowing us to find clues of preslip on the nearby Futagawa FZ, which were not found by the latter study. The Kumamoto case might not be a good application of the FTLS.

The picture changes markedly after the $M$7.3 mainshock: the $b$-values in the Kumamoto source area that largely fluctuated during the first several months (Fig. 3c), increased significantly, compared with the background (also see section 2 and Nanjo et al., 2019). This is consistent with the findings by Gulia and Wiemer (2019), who suggested that the chance of a subsequent large event was substantially small, although the Kumamoto aftershock sequence included many small events. No subsequent large event has occurred so far.

The FTLS system is quite new, and some cases, except for the Kumamoto one, would provide support for it. We agree with Gulia et al. (2020) that it is too early to use the current version of the FTLS routinely for making decisions about civil protection and public communications. More extensive tests of sensitivity and robustness are needed, as pointed out by those authors. Based on our studies, it would be very worthwhile to improve the FTLS in ways that would allow spatial variability of *b*-values to be taken into consideration, and if an indication of preslip appeared, this would trigger a red traffic light. Also, the problem may be a lack of another type of sensitivity, namely only 3 levels of sensitivity, red, yellow and green. Perhaps a wider range of sensitivities would be more useful, but then the name would likely have to change to something other than the FTLS, since traffic lights do not have more than 3 colors.

5. **Conclusion**

This paper reviews previous studies on spatial and temporal patterns of seismicity in the Kumamoto region, in Japan, where some large earthquakes took place in 2016, including one with *M*7.3 (Nanjo et al., 2016; 2019; Nanjo and Yoshida, 2017). Based on stress-dependent laws of statistical seismology, the stress evolution during this earthquake sequence could be illustrated.

Based on the outcomes, we provide additional insight into the FTLS (Gulia and Wiemer, 2019).

Using the $b$-value of the GR frequency-magnitude law and the $p$-value of the OU aftershock-decay law, it was shown that a zone of low $b$-values (indicative of high stress) lay near the eventual epicenters of the $M$6.5 and $M$7.3 earthquakes before the start of the Kumamoto sequence, and that after it, the overall trend in stress decreased along the Futagawa-Hinagu FZ. Detailed analysis suggested aseismic preslips and afterslips along the causative FZ. Stress reduction just prior to the $M$7.3 mainshock near its epicenter was suggested to be locally induced by preslips. Based on the review, we noted that the observed $b$-values were indicative of a green FTLS status before the $M$7.3 event, in contrast to a red FTLS status shown by Gulia and Wiemer (2019), who did not study spatial variability of $b$-values to help detect aseismic preslips that occurred locally on the Futagawa FZ. In our view, taking this $b$-value variability into account, together with using other seismological and geological information such as $p$-values and the spatial configuration of active faults, is important to improve the FTLS. It would also be of interest for future work to improve FTLS, taking into account the triggering of a red traffic light when signals to suggest ongoing preslips are detected.

**CRediT authorship contribution statement**

**K. Z. Nanjo:** Conceptualization, Funding acquisition, Formal analysis, Methodology,

Software, Writing -original draft, Writing -review & editing. **J. Izutsu:** Conceptualization, Funding acquisition, Writing -review & editing. **Y. Orihara:** Conceptualization, Writing -review & editing. **M. Kamogawa:** Conceptualization, Writing -review & editing.

**Declaration of competing interests**

The authors declare that they have no known competing financial interests or personal relationships that could have appeared to influence the work reported in this paper.

**Acknowledgements**

The authors thank the Editor (L. Chen), P. Han, C.-H. Chan, and an anonymous reviewer for their valuable comments. This study was partially supported by the Ministry of Education, Culture, Sports, Science and Technology (MEXT) of Japan, under its The Second Earthquake and Volcano Hazards Observation and Research Program (Earthquake and Volcano Hazard Reduction Research) (K.Z.N., J.Z., Y.O., and M.K.) and under it STAR-E (Seismology TowArd Research innovation with data of Earthquake) Program Grant Number JPJ010217 (K.Z.N.), JSPS KAKENHI Grant Number JP 20K05050 (K.Z.N., J.Z., and M.K.), the Collaboration Research Program of IDEAS, Chubu University IDEAS202111 (K.Z.N., J.Z., and M.K.), Tokio Marine Kagami Memorial Foundation (K.Z.N.), and the Chubu Electric Power's research based on selected proposals (K.Z.N.). The seismicity analysis software package ZMAP (Wiemer, 2001), used for Figs. 2, 3a-c, and 4, was obtained from http://www.seismo.ethz.ch/en/research-and-teaching/products-software/software/ZMAP. The Generic Mapping Tools (Wessel et al., 2013), used for Figs. 1, 2, and 3d, are an open-source collection (https://www.generic-mapping-tools.org).

**Data availability**

The JMA earthquake catalog used in this study was obtained from http://www.data.jma.go.jp/svd/eqev/data/bulletin/index_e.html. We obtained the afterslip planes shown in Fig. 2 from those shown in Fig. 1 of Pollitz et al. (2017).

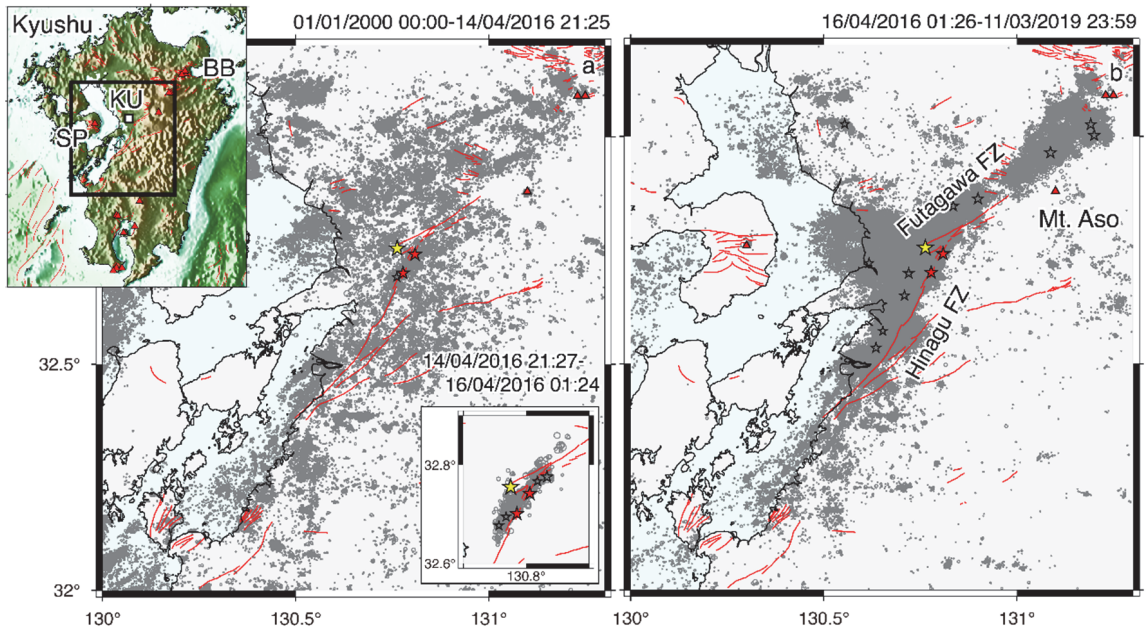

Fig. 1. The 2016 Kumamoto earthquake sequence ($M \geq 0$). (a) Map of earthquakes before the start of the Kumamoto earthquakes (January 1, 2000 to April 14, 2016, 20:15). Stars show epicenters of the $M$6.5 and $M$6.4 foreshocks (red) and $M$7.3 mainshock (yellow), and $M$5 event on June 8, 2000 (open). Triangles indicate volcanos. Red lines indicate mapped faults. Inset at the upper-left corner shows the study region (black rectangle). White square indicates the City of Kumamoto (KU). BB and SP represent Beppu Bay and Shimabara Peninsula, respectively. Inset at the lower-right corner shows a map during the period from the time of the $M$6.5 foreshock to the time of the $M$7.3 mainshock. Stars indicate $M \geq 5$ events during this period. (b) Same as (a) for aftershocks (April 16, 2016, 1:25 to March 11, 2019, 23:59). Stars indicate $M \geq 5$ events in this period.

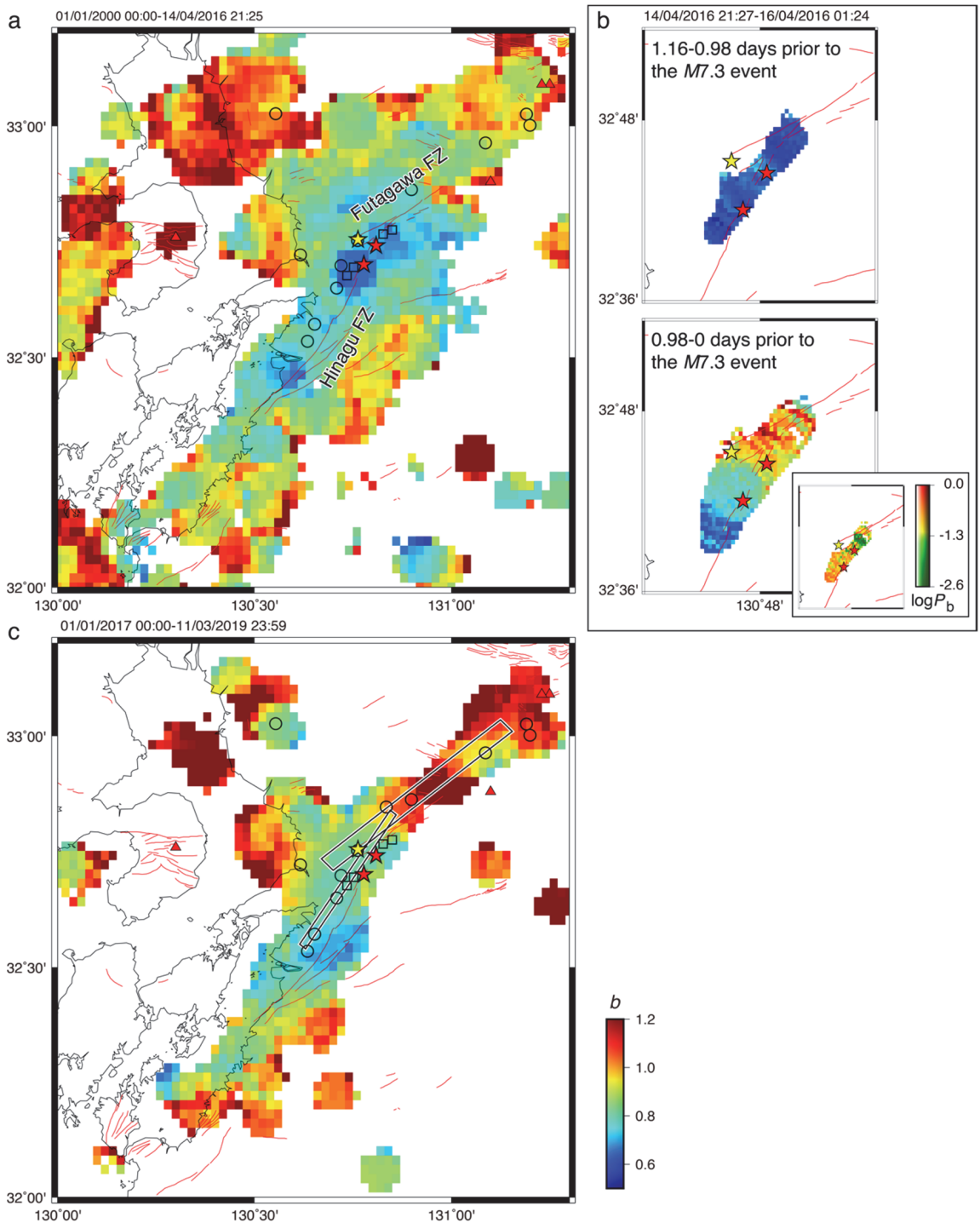

Fig. 2. Map of *b*-values. (a) *b*-values were obtained from events before the start of the Kumamoto earthquake sequence (from January 1, 2000 to immediately before the *M*6.5 foreshock). Single *b*-values were calculated for events falling in a cylindrical volume with radius *r* = 5 km, centered at each node on a 0.02º × 0.02º grid (modified from Nanjo et al., 2016). Overlapped are the epicenters of the *M*7.3 mainshock (yellow star), *M*6.5-class foreshocks (red stars), *M* ≥ 5 foreshocks (squares) since January 1, 2000, and *M* ≥ 5 aftershocks (circles) until March 11, 2019. Triangles and red lines are the same as in Fig. 1. (b) Results of *b*-value analysis for the period from the *M*6.5 foreshock to the *M*7.3 mainshock (modified from Nanjo and Yoshida, 2017). Top and bottom panels: periods 1.16-0.98 and 0.98-0 days prior to the *M*7.3 mainshock, respectively. The mapping procedure is the same as that for (a), except for a smaller node spacing (0.005° × 0.005°). Inset: log$P_b$ for comparison in *b* between the periods 1.16-0.98 (top panel) and 0.98-0 (bottom panel) days. (c) Same as (a) for the period after the *M*7.3 mainshock (from January 1, 2017 to March 11, 2019) (modified from Nanjo et al., 2019). Superimposed are after-slip planes (black rectangles) used by Pollitz et al. (2017).

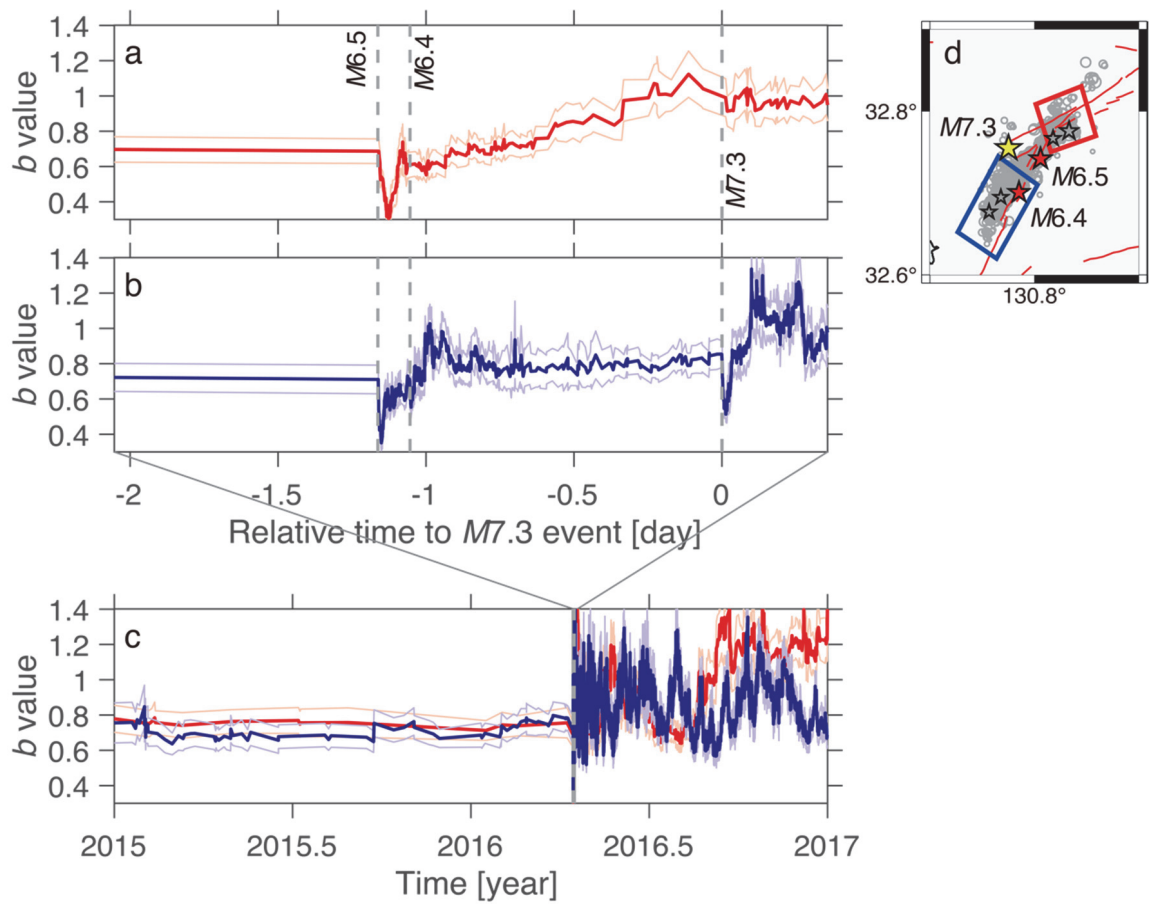

Fig. 3. *b*-value timeseries. (a) Plot of *b*-values as a function of relative time to the *M*7.3 quake. Events in the region in red in (d) were used. Uncertainties are according to Schorlemmer et al. (2003). Vertical dashed lines indicate moments of the *M*6.5, *M*6.4, and *M*7.3 earthquakes. (b) Same as (a) for the region in blue in (d). (c) The plotting procedure was the same as that for (a) and (b), except for a longer period (2015-2017 years). In this plot, vertical dashed lines indicating moments of the *M*6.5, *M*6.4, and *M*7.3 earthquakes overlap with each other. (d) Map to indicate the two regions in red and blue in which events used for the *b*-value timeseries analysis in (a)-(c) are included. For other symbols, see Fig. 1.

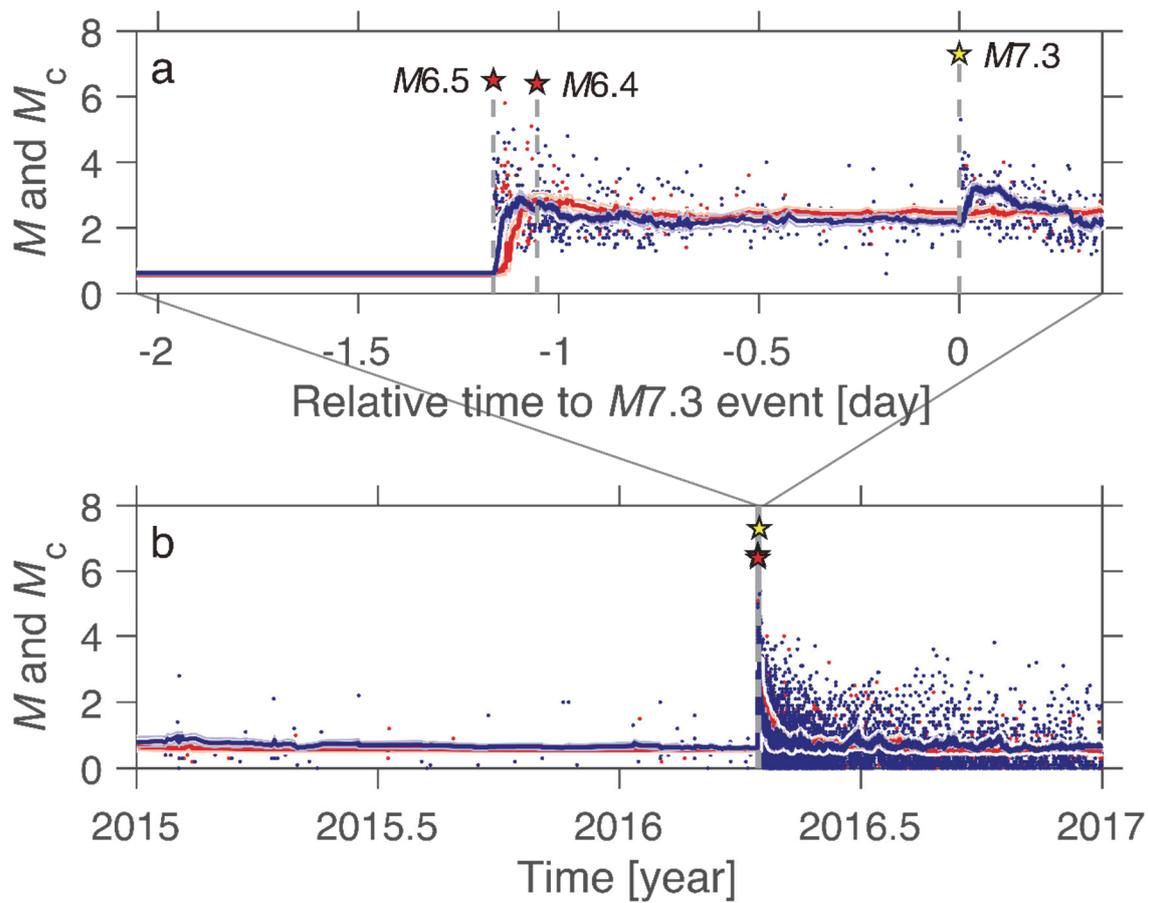

Fig. 4. Timeseries of $M_c$. (a) Same as Figs. 3a,b for $M_c$, where uncertainties are according to Woessner et al. (2005). Included in (a) is the time dependence of $M$ for events used to compute $M_c$. These events were also used for computing the $b$-values in Fig. 3 Vertical dashed lines indicate moments of earthquakes with $M$6.5 (red star), $M$6.4 (red star), and $M$7.3 (yellow star). (b) Same as (a) except for a longer period (2015-2017). $M_c$-values that had been about 1 before the $M$6.5 earthquake, consistent with previous studies (Nanjo et al., 2010; Schorlemmer et al., 2018). Immediately after the start of the Kumamoto earthquake, $M_c$ increased to about 3, confirming the result of Nanjo and Yoshida (2017), and then showed a gradually decreasing trend to the background level ($M_c \sim 1$). A zoom-in figure showing $M_c$-values around the period between the $M$6.5 and $M$7.3

events is given in (a).